\documentclass[acmsmall,screen,normalem]{acmart}
\usepackage{amsthm}
\usepackage[normalem]{ulem}
\newtheorem{theorem}{Theorem}
\newtheorem{definition}{Definition}

\newtheorem{assumption}{Assumption}

\title{A Trace-based Approach for Code Safety Analysis}
\author{Hui Xu}
\affiliation{%
  \institution{Fudan University}
  \country{China}
}
\date{October 2025}
\begin{abstract}
Rust is a memory-safe programming language that disallows undefined behavior. Its safety guarantees have been extensively examined by the community through empirical studies, which has led to its remarkable success. However, unsafe code remains a critical concern in Rust. By reviewing Rust’s safety design and analyzing real-world Rust projects, this paper establishes a systematic framework for understanding unsafe code and undefined behavior, and summarizes the soundness criteria for Rust code. It further derives actionable guidance for achieving sound encapsulation.
\end{abstract}

\begin{document}

\maketitle

\section{Introduction}
This paper is based on Rust, a system programming language that emphasizes memory safety. Starting from the safety promise of Rust, \textit{i.e.,} safe code cannot cause undefined behavior~\cite{safety_promise}, we first establish a main theorem about the relationship between unsafe code and undefined behavior. \uline{Informally, undefined behavior originates exclusively from unsafe code and is solely determined by the safety constraints of that unsafe code.}

The theorem is derived from an observation about unsafe code in Rust: all unsafe functions must declare safety contracts that need to be upheld to prevent undefined behavior. These contracts serve as sufficient conditions and are applicable to all usage scenarios.
The theorem enables a \textit{trace-based approach} to safety encapsulation. This approach is analogous to taint analysis, where unsafe code within a function can be viewed as the taint source and the function entrance/exits as the sinks. Under this perspective, all propagations of undefined behavior risks introduced by unsafe code must not cross encapsulation boundaries. By treating the safety contract of a safe function as empty, we can derive a corollary that specifies how to achieve safety encapsulation for both safe and unsafe functions. \uline{Informally, a free-standing function must ensure the safety contracts of its unsafe callees are upheld if the safety contracts of itself can be upheld.}

We further extend the application of the theorem to structs, which consist of data fields and associated functions. Structs are more complex because their associated functions are interdependent, operating on a shared internal state. As a result, it is inappropriate to consider the soundness of a single associated function in isolation. To decouple the relationship, we find it is useful to introduce safety invariants for struct types, which specify the properties that all valid instances of a struct must uphold. If any associated function of a struct may introduce an invalid instance that violates the safety invariant, that function should be declared unsafe. In this way, other functions of the struct can rely on the type invariant to achieve sound encapsulation. \uline{Informally, a struct must ensure that, for each of its associated functions, if the function’s own safety contract holds, then the struct’s safety invariant is preserved and the safety contracts of any unsafe callees are upheld.}

The remainder of this paper is organized as follows. We first introduce the notations used, followed by the formal definition and proof of the main theorem. We then discuss how to achieve sound encapsulation for both safe and unsafe functions, and subsequently extend the analysis to structs and modules. Finally, we discuss potential applications of this work and conclude.

\section{Notations}
\begin{description}
  \item[$c$:] a constructor.
  \item[$f$:] a free function.
  \item[$f_s$:] a safe free function.
  \item[$f_u$:] an unsafe free function.
  \item[$i$:] an instance.
  \item[$lc$:] a literal constructor.
  \item[$m$:] a struct method with a \texttt{self}, \texttt{\&self}, or \texttt{\&mut self} receiver.
  \item[$p$:] a projection.
  \item[$C$:] a set of associated functions that return \texttt{Self} or a type that wraps \texttt{Self}.
  \item[$F$:] a set of functions.  
  \item[$Fn$:] a set of associated functions. 
  \item[$Fp$:] a set of field projections.
  \item[$M$:] a set of associated methods in a struct.
  \item[$S$:] a set of structs.
  \item[$SC$:] a safety contract.
  \item[$SI$:] safety invariants.
  \item[$UB$:] undefined behavior.
  \item[$UC$:] unsafe code.
  \item[$V$:] a set of variables.
  \item[$\mathcal{P}$:] a program.
  \item[$\mathcal{P}_{f_s}$:] a program that calls $f_s$ and contains no unsafe code.
  \item[$\mathcal{P}_{f_u}$:] a program that calls $f_u$ and contains no other unsafe code.
  \item[$\mathcal{S}$:] a struct.
  \item[$\mathbb{M}$:] a module.
\end{description}

\section{Main Theorem}
In this section, we establish the main theorem, which characterizes the relationship between undefined behavior and unsafe code in Rust.

\begin{theorem}[Main Theorem]\label{theorem:main}
Let $\mathcal{P}$ be a well-typed Rust program. Then, undefined behavior can occur in $\mathcal{P}$ only if $\mathcal{P}$ contains unsafe code and $\mathcal{P}$ violates its associated safety contract. Formally,
\[
  \mathcal{P} \Downarrow UB 
  \; \Rightarrow \; 
  \mathcal{P} \supseteq UC \, \wedge\, \mathcal{P} \not\models SC_{UC},
\]
where $UC$ denotes a piece of unsafe code (typically an unsafe function call) contained in $\mathcal{P}$, and $SC_{UC}$ denotes the safety contract for using the unsafe code.

\end{theorem}

This theorem asserts that undefined behavior (UB) originates exclusively from unsafe code and is entirely determined by the safety contract of the corresponding unsafe code.

The proof of this theorem is based on the soundness criteria for safe code (Definition~\ref{def:criterion_safe}) and unsafe code (Definition~\ref{def:criterion_unsafe}), which are defined later in this chapter.
For simplicity, we assume that unsafe code consists solely of calls to unsafe functions.
This assumption does not restrict generality, as the same reasoning applies to other forms of unsafe operations, such as raw pointer dereferences or accesses to static mutable variables, which likewise possess well-defined safety contracts.

\subsection{Soundness Criterion of Safe Code}

Rust’s safety promise is originally stated as follows~\cite{safety_promise}:
\begin{quote}
\itshape
“For Rust, this (soundness) means well-typed programs cannot cause Undefined Behavior. This promise only extends to safe code however; for unsafe code, it is up to the programmer to uphold this contract.”
\end{quote}

Assuming the compiler is sound, only well-typed code can pass type checking. Consequently, the safety promise can be formally expressed as follows:

\begin{definition}[Safety Promise of Rust]\label{def:safety_promise}
Safe code cannot cause undefined behavior, and only unsafe code may exhibit undefined behavior. Formally,
\[
  \forall \mathcal{P}, \; \mathcal{P} \not\supseteq UC\; \Rightarrow\; \mathcal{P} \not\Downarrow UB.
\]
\end{definition}

Building upon the safety promise, we formalize the soundness criterion for safe Rust code as follows.

\begin{definition}[Soundness criterion of Safe Functions]\label{def:criterion_safe}
A safe function $f_s$, regardless of whether it contains internal unsafe code, is \emph{sound} if and only if
\[
  \forall \mathcal{P},\;
  \mathcal{P} \not\supseteq UC\,\wedge\, f_s \in \mathrm{Callee}(\mathcal{P})  
  \;\Rightarrow\;
  \mathcal{P} \not\Downarrow UB,
\]
which can be simplified as
\[
  \forall \mathcal{P}_{f_s}, \; \mathcal{P}_{f_s} \not\Downarrow UB,
\]
where $\mathcal{P}_{f_s}$ denotes a program that calls $f_s$ and contains no unsafe code, including unsafe code introduced by other safe functions with internal unsafe implementations, unless those functions also satisfy this soundness criterion.
\end{definition}

\subsection{Soundness Criterion of Unsafe Code}
Based on best practices observed in real-world Rust projects, particularly in the standard library, we find that unsafe functions are typically accompanied by well-documented safety requirements or contracts. These contracts exhibit two essential properties:

\begin{itemize}
\item \textit{Pervasiveness:} Every unsafe function is associated with a safety contract that must be upheld to prevent undefined behavior. Such contracts serves as sufficient conditions for safety.
\item \textit{Uniformity:} The safety contract associated with a given function is consistent across all of its call sites.
\end{itemize}

Building upon these observations, we make the following formal assumption about unsafe functions and their safety contracts.
\begin{assumption}[Properties of Safety Contracts for Unsafe Functions]\label{assumption:safety_contraint}
\[
  \forall f_u, \; \exists SC_{f_u} \; \text{s.t.} \;
  \forall \mathcal{P}_{f_u}, \; \mathcal{P}_{f_u} \models SC_{f_u} \;\Rightarrow\; \mathcal{P}_{f_u} \not\Downarrow UB,
\]
where $SC_{f_u}$ denotes the safety contract of an unsafe function $f_u$, and $\mathcal{P}_{f_u}$ is a program that uses $f_u$ and contains no other unsafe code, including unsafe code introduced by other APIs with internal unsafe implementations, unless those APIs satisfy Definition~\ref{def:criterion_safe}.
\end{assumption}

Based on Assumption~\ref{assumption:safety_contraint}, we can derive the following soundness criterion for unsafe functions:
\begin{definition}[Soundness Criterion of Unsafe Functions]\label{def:criterion_unsafe}
An unsafe function $f_u$ associated with safety contract $SC_{f_u}$ is sound if and only if
\[
  \forall \mathcal{P}_{f_u}, \;
  \mathcal{P}_{f_u} \models SC_{f_u} \;\Rightarrow\; \mathcal{P}_{f_u} \not\Downarrow UB.
\]
\end{definition}

With this criterion, we can now prove the correctness of Theorem~\ref{theorem:main}. Specifically, Definition~\ref{def:safety_promise} implies that $\mathcal{P} \supseteq UC$ is a necessary condition for undefined behavior, while Definition~\ref{def:criterion_unsafe} implies that $\mathcal{P}_{f_u} \not\models SC_{f_u}$ is also a necessary condition.

\section{Sound Function Encapsulation}

From Theorem~\ref{theorem:main}, we can derive corollaries for achieving sound encapsulation of both safe and unsafe functions.

\begin{corollary}[Safe Function Encapsulation]\label{corollary:safe_function}
A safe function $f_s$ is sound if either it contains no unsafe code, or the safety contracts of all unsafe callees are upheld. Formally,
\[
  \forall f'_u \in \mathrm{UnsafeCallee}(f_s), \; f_s \models SC_{f'_u},
\]
where $\mathrm{UnsafeCallee}(f_s)$ denotes the set of unsafe functions called by $f_s$, 
and $SC_{f_u}$ denotes the safety contract of $f_u$.
\end{corollary}

\begin{corollary}[Unsafe Function Encapsulation]\label{corollary:unsafe_function}
An unsafe function $f_u$ is sound if the safety contracts of all its unsafe callees are upheld. Formally,
\[
  \forall f'_u \in \mathrm{UnsafeCallee}(f_u), \;
  f_u \cup SC_{f_u} \models SC_{f'_u}.
\]
\end{corollary}

We can unify Corollary~\ref{corollary:safe_function} and Corollary~\ref{corollary:unsafe_function} by treating a safe function as having an empty constraint set. In this way, we obtain the following general corollary for function encapsulation.

\begin{corollary}[Sound Function Encapsulation]\label{corollary:function_encapsulation}
A function $f$ (safe or unsafe) is sound if the safety constraints of all its unsafe callees are satisfied. Formally,
\[
  \forall f_u \in \mathrm{UnsafeCallee}(f), \;
  f \cup SC_{f} \models SC_{f_u}.
\]
\end{corollary}

Note that although the soundness criteria in Definition~\ref{def:criterion_safe} and Definition~\ref{def:criterion_unsafe} are recursive, the encapsulation rule defined in Corollary~\ref{corollary:function_encapsulation} is not. In the presence of recursive cases, \textit{e.g.,} \texttt{f} $\rightarrow$ \texttt{f}, Corollary~\ref{corollary:function_encapsulation} remains applicable by assuming that the safety contracts of all functions involved in the recursion are declared, irrespective of whether they are sound. Developers can then directly apply their safety contracts and detect contradictions via fixed-point iteration.

\section{Struct Soundness}
This section begins by defining the setting of structs in Rust and then extends the previous results to this setting.

\subsection{Settings of Structs}
Following the official Rust book\cite{rust_struct}, a Rust struct can be characterized by the following components:
\[
\mathcal{S} = \{Fd, lc, Fp, Fn\}.
\]

\begin{itemize}
    \item $\mathit{Fd}$ is a collection of fields that constitute the internal state of the struct;
    \item $lc$ is the struct literal used to construct or initialize instances of the struct;
    \item $Fp$ are field projections that directly access the fields of an instance, \textit{e.g.,} \texttt{s.a};
    \item $Fn$ are associated free functions defined in an \texttt{impl} block.
\end{itemize}

Structs are more complex because the associated functions of a struct are interdependent, making it generally impossible to consider each function separately. For example, functions with a \texttt{\& self} receiver can only be invoked on constructed instances; functions with a \texttt{\&mut self} receiver may mutate the internal state of an instance, thereby affecting the safety encapsulation of other associated functions. 

To aid reasoning, we further separate associated functions $Fn$ into three types: 
\[
Fn = \{C, F, M\}.
\]
\begin{itemize}
    \item $C$ are associated functions that return \texttt{Self} or a type that wraps \texttt{Self} (e.g., \texttt{Box<Self>}). We also treat such functions as constructors. This modeling choice extends the default notion of constructors in Rust, which includes only literal constructors~\cite{rust_constructors}. In contrast, free-standing functions that return a struct instance are not considered constructors, as they internally rely on existing struct constructors. 
    \item $M$ are \textit{methods}, \textit{i.e.,} associated functions with a \texttt{self}, \texttt{\&self} or \texttt{\&mut self} receiver, including both inherent methods and trait methods.
    \item $F$ are the remaining associated functions, \textit{i.e.,} $F = Fn \setminus C \setminus M$.
\end{itemize}

To decouple these dependencies and facilitate reasoning, we further rely on the safety invariant of a struct type, which characterizes the contract that all valid instances must uphold~\cite{safety_invariant}. Safety invariants have become a de facto standard in Rust-for-Linux.

\begin{assumption}[Safety Invariants of Struct]\label{assumption:struct_invariant}
For each struct type $\mathcal{S}$, there may exist a safety invariant $SI_\mathcal{S}$ such that, if defined,
\[
  \forall i \in \mathrm{Instances}(\mathcal{S}),\;
  i \in \mathrm{ValidInstances}(\mathcal{S})
  \;\iff\;
  i \models SI_\mathcal{S},
\]

where $\mathrm{ValidInstances}(\mathcal{S})$ is the set of instances for which any program using $i$ safely does not lead to undefined behavior:
\[
  \forall i \in \mathrm{ValidInstances}(\mathcal{S}), \; \forall \mathcal{P}_i, \; \mathcal{P}_i \not\Downarrow UB.
\]
\end{assumption}

\subsection{Soundness Criteria of Structs}
In Rust, each field is by default visible at the module level rather than the struct level, which in turn affects the availability of struct literals and field projections outside the defining module. To align with module-level soundness, we do not need to consider undefined behavior arising from literal constructors and field projections used within the defining module. Based on this observation, we define the following soundness criteria for structs.

\begin{definition}[Soundness Criteria of Structs]\label{def:criteria_struct}
A struct
\(
\mathcal{S} = \{Fd, lc, Fp, C, F, M\}
\)
with safety invariant $SI_{\mathcal{S}}$, defined in module $\mathbb{M}$, is \emph{sound} if and only if the following conditions hold:

\begin{enumerate}
  \item If the literal constructor $lc$ is available outside the module defining the struct:
  \[
    \forall \mathcal{P}_{lc} \notin \mathbb{M},\;
    \mathcal{P}_{lc} \Downarrow i
    \;\Rightarrow\;
    i \models SI_{\mathcal{S}} \;\wedge\; \mathcal{P}_{lc} \not\Downarrow UB.
  \]
  where $i$ is the struct instance created by $lc$.

  \item For all field projections available outside the module defining the struct:
    \[
      \forall p \in Fp,\;
      \forall \mathcal{P}_{p}(i) \notin \mathbb{M},\;
      i \models SI_{\mathcal{S}}
      \;\wedge\;
      \mathcal{P}_{p}(i) \Downarrow i'
      \;\Rightarrow\;
      i' \models SI_{\mathcal{S}} \;\wedge\; \mathcal{P}_{p}(i) \not\Downarrow UB,
    \]
where $i$ is the initial struct instance and $i'$ is the instance resulting from executing $\mathcal{P}_{p}(i)$.

  \item For all constructors excluding the literal constructor:
  \[
    \forall c \in C,\;
    \forall \mathcal{P}_{c},\;
    \mathcal{P}_{c} \models SC_c
    \;\wedge\;
    \mathcal{P}_{c} \Downarrow i
    \;\Rightarrow\;
    i \models SI_{\mathcal{S}} \;\wedge\; \mathcal{P}_{c} \not\Downarrow UB.
  \]

  \item For all methods:
  \[
    \forall m \in M,\;
    \forall \mathcal{P}_{m}(i),\;
    i \models SI_{\mathcal{S}} \;\wedge\;
    \mathcal{P}_{m}(i) \models SC_m
    \;\wedge\;
    \mathcal{P}_{m}(i) \Downarrow i'
    \;\Rightarrow\;
    i' \models SI_{\mathcal{S}} \;\wedge\; \mathcal{P}_{m}(i) \not\Downarrow UB.
  \]

  \item For the remaining associated functions:
  \[
    \forall f \in F,\;
    \forall \mathcal{P}_f,\;
    \mathcal{P}_f \models SC_f
    \;\Rightarrow\;
    \mathcal{P}_f \not\Downarrow UB.
  \]
\end{enumerate}
\end{definition}

Note that we treat the destructor in the same way as other methods because it also requires the \texttt{\&mut self} argument. We do not need to consider the destructor separately as it is automatically added by the compiler and is safe. Although developers can implement the \texttt{Drop} trait with unsafe code, such unsafe code should be properly encapsulated within the \texttt{drop()} function, which is a safe function and has no safety contract. 

\subsection{Sound Struct Encapsulation}

For each item in the soundness criteria of Definition~\ref{def:criteria_struct}, we define a corresponding encapsulation rule in Corollary~\ref{corollary:struct_encapsulation}.

\begin{corollary}[Sound Struct Encapsulation]\label{corollary:struct_encapsulation}
A struct
\(
\mathcal{S} = \{Fd, lc, Fp, C, F, M\}
\)
with safety invariant $SI_{\mathcal{S}}$, defined in module $\mathbb{M}$, is sound if it satisfies the following encapsulation rules:
\begin{enumerate}
  \item \textit{Literal constructor availability}:
  if the safety invariant $SI_\mathcal{S}$ is non-empty, then the literal constructor
  $lc$ must not be available to programs outside $\mathbb{M}$, i.e.,
  \[
    SI_\mathcal{S} \neq \emptyset
    \;\Rightarrow\;
    \forall \mathcal{P} \notin \mathbb{M},\;
    lc \notin \mathrm{available}(\mathcal{P}).
  \]

  \item \textit{Field projection availability}:
  all field projections available outside $\mathbb{M}$ must preserve the safety invariant:
  \[
    \forall p \in Fp,\;
    \forall \mathcal{P}_{p}(i) \notin \mathbb{M},\;
    i \models SI_{\mathcal{S}}
    \;\wedge\;
    \mathcal{P}_{p}(i) \Downarrow i'
    \;\Rightarrow\;
    i' \models SI_{\mathcal{S}}.
  \]

  \item \textit{Sound constructor encapsulation}:
  a constructor $c\in C$ is sound if, assuming its safety contract $SC_c$ holds, all struct instances constructed by $c$ satisfy the safety invariant of the struct, and the safety contracts of all its unsafe callees are upheld. Formally,
  \[
    \Big(
      \forall \mathcal{P}_{c},\;
      \mathcal{P}_{c} \models SC_c
      \;\wedge\;
      \mathcal{P}_{c} \Downarrow i
      \;\Rightarrow\;
      i \models SI_{\mathcal{S}}
    \Big)
    \;\wedge\;
    \forall f_u \in \mathrm{UnsafeCallee}(c),\;
    c \cup SC_c \models SC_{f_u}.
  \]
  The first part ensures that the constructor does not generate invalid instances that may affect subsequent usage; the second part ensures that the constructor itself does not lead to undefined behavior due to its internal unsafe code.

  \item \textit{Sound method encapsulation}:
  an associated method $m\in M$ is sound if, assuming its safety contract $SC_m$ holds, all struct instances modified by $m$ satisfy the safety invariant of the struct, and the safety contracts of all its unsafe callees are upheld. Formally,
    {\small
    \[
    \Big(
      \forall \mathcal{P}_{m}(i),\!
      i \models SI_{\mathcal{S}}
      \!\wedge\!
      \mathcal{P}_{m}(i) \models SC_m
      \!\wedge\!
      \mathcal{P}_{m}(i) \Downarrow i'
      \!\Rightarrow\!
      i' \models SI_{\mathcal{S}}
    \Big) \!\wedge\!
    \forall f_u \in \mathrm{UnsafeCallee}(m),\!
    m \cup SC_m \models SC_{f_u}.
    \]
    }

  \item \textit{Remaining free function encapsulation}:
    For an associated free function $f \in F$, assuming its safety contract $SC_f$ holds, the safety contracts of all its unsafe callees are upheld. Formally,
    \[
      \forall f_u \in \mathrm{UnsafeCallee}(f),\;
      f \cup SC_f \models SC_{f_u}.
    \]
\end{enumerate}
\end{corollary}

\subsection{Considerations of Traits}
This work treats traits as a set of methods. When a struct implements a trait, the trait’s methods are added to the struct, so the previous results still hold. 

In addition, some traits are unsafe and have their own safety contracts for implementation. Such implementation issues are independent of struct soundness.

\begin{assumption}[Properties of Safety Contracts for Unsafe Traits]\label{assumption:unsafe_trait_contract}
For each unsafe trait $T$, there exist a safety contract $SC_{T}$ such that, for any implementation $\mathrm{Impl}(T)$ of the trait,
\[
  \mathrm{Impl}(T) \models SC_{T} \;\Rightarrow\; 
  \big( \forall f \in \mathrm{Impl}(T), \; \forall \mathcal{P}_f, \; 
      \mathcal{P}_f \models SC_f \;\Rightarrow\; \mathcal{P}_f \not\Downarrow UB \big).
\]
\end{assumption}

This assumption can similarly be extended to free functions and methods; we omit the details for brevity.

\section{Module and Crate Soundness}
In Rust, software is organized into crates, each of which may consist of multiple modules. 
A crate is considered sound if all of its constituent modules are sound. 
Thus, it suffices to characterize the soundness of individual modules. 
Each module consists of a collection of static variables $V$, free-standing functions $F$, and structs $S$.

\begin{definition}[Module Soundness Criteria]\label{def:criteria_module}
A module 
\(
\mathbb{M} = \{V, F, S\}
\)
is sound if and only if both of the following conditions hold:

\begin{enumerate}
  \item All public free-standing functions are sound:
  \[
    \forall f \in F,\;\forall \mathcal{P}_f \not\in \mathbb{M}, \;
    \mathcal{P}_f \models SC_f \;\Rightarrow\; \mathcal{P}_f \not\Downarrow UB
  \]
  
  \item  All public structs are sound:
  \[
    \forall s \in S,\;\forall \mathcal{P}_s \not\in \mathbb{M}, \;
    \mathcal{P}_s \models SC_s \;\Rightarrow\; \mathcal{P}_s \not\Downarrow UB
  \]
\end{enumerate}
\end{definition}

The soundness criteria consider only the public components of a module and decouple the requirements for functions and structs. Consequently, we can directly refer to Corollary~\ref{corollary:function_encapsulation} and Corollary~\ref{corollary:struct_encapsulation} to achieve sound module encapsulation.

Note that we do not need to consider static variables or interleavings among functions and structs, even if they share some static mutable variables. These static mutable variables are distinct from method fields: accessing method fields is safe, whereas accessing static mutable variables is unsafe. The safety contracts for accessing such static mutable variables are assumed to be enforced by the corresponding functions and structs, ensuring that these functions and structs are themselves sound. For static variables with internal mutability, our approach assumes that the safety contracts of functions or the safety invariants of structs account for any potential undefined behavior arising from internal mutation. Consequently, as long as each individual function and struct of a module is sound, the entire module is considered sound.

\section{Application Discussion}
The soundness criteria and encapsulation approach may serve as practical guidelines for real-world Rust software development. They highlight the necessity of safety contracts in soundness reasoning and specify the properties that such contracts should satisfy. Developers can follow this approach to achieve sound encapsulation with clearly documented safety contracts for functions and structs. This approach is also useful for identifying which component is unsound when a downstream crate relying on another crate exhibits undefined behavior. Moreover, program analysis tools may adopt this reasoning framework for soundness verification.

\section{Conclusion}
This work introduces a trace-based approach for analyzing unsafe code and verifying soundness. The approach is grounded in a main theorem, which states that undefined behavior originates exclusively from unsafe code and is solely determined by the safety contracts of that code.
We prove the correctness of this theorem under assumptions on the properties of safety contracts.
Building on this theorem, we have derived soundness criteria for both functions and structs, along with guidance for achieving sound encapsulation.

\bibliographystyle{ACM-Reference-Format}
\bibliography{ref}

\end{document}